\documentclass[12pt]{iopart}

\usepackage{cite}

\def\vep{\varepsilon}

\def\be{\begin{equation}}
\def\ee{\end{equation}}
\def\bea{\begin{eqnarray}}
\def\eea{\end{eqnarray}}
\newcommand{\beqn}{\begin{eqnarray}}
\newcommand{\eeqn}{\end{eqnarray}}
\newcommand{\beqnn}{\begin{eqnarray*}}
\newcommand{\eeqnn}{\end{eqnarray*}}

\begin{document}

\title{Asymptotical photon distributions in the dissipative Dynamical Casimir Effect}

\author{V V Dodonov}

\address{
$\mbox{}^1$ Instituto de F\'{\i}sica, Universidade de Bras\'{\i}lia,
Caixa Postal 04455, 70910-900 Bras\'{\i}lia, DF, Brazil }

\eads{\mailto{vdodonov@fis.unb.br} }

\begin{abstract}

Asymptotical formulas for the photon distribution function of a quantum oscillator
with time-dependent frequency and damping coefficients, interacting with a thermal reservoir,
are derived in the case of a large mean number of quanta. 
Different regimes of excitation of an initial thermal state with an arbitrary temperature are considered.
New formulas are used to predict the 
statistical properties of the electromagnetic field created in the experiments on the Dynamical Casimir Effect
which are now under preparation. 

\end{abstract}


\section{Introduction}

The so-called {\em Dynamical Casimir Effect\/} (DCE), i.e., a
generation of photons from vacuum due to the motion of neutral boundaries,
was a subject of numerous theoretical studies for almost 40 years
(for the specific reviews on the DCE see \cite{D-rev1,DD-rev2,DCas60}; general reviews on 
the Casimir physics, including
different manifestations of the static and dynamical Casimir effects, can be found, e.g., in \cite{Milton04,Most09}).
One of the most important results obtained during a decade before and after 2000 
was a prediction of a possibility of an observation of the DCE
in a laboratory, using high-$Q$ cavities with dimensions of the order of few centimeters
and resonance frequencies of the order of few GHz or higher. Namely, calculations performed by several groups of authors 
\cite{D95,DK96,Lamb,Plun,Croc1,Croc04,Plun04}
(using quite different approaches) resulted in the same conclusion: if one
could arrange {\em periodical\/} changes of parameters of some cavity
(its dimensions or properties of the walls) for a sufficiently long time,
then initial vacuum or thermal fluctuations of the electromagnetic field 
could be amplified to a detectable level due to the effect of 
{\em parametric resonance\/}. 
This result stimulated the work of several experimental groups, from which
the MIR group of the university of Padua \cite{Pad-Cas60} seems to be close to
a success (the name of the experiment MIR means `Motion Induced Radiation'; 
this term was coined in \cite{Lamb}). 
Therefore it seems interesting to calculate the statistical properties of
quantum states which could be obtained under realistic experimental
conditions.

The simplest model of the DCE in cavities is that of a quantum oscillator
with a time-dependent frequency,
describing the selected field mode which is in resonance with time variations
of the cavity parameters. Such a model was first proposed in \cite{Man91,Sas} 
and later it was developed in \cite{D95,DK96}. 
In the ideal case (without losses or
interactions with other degrees of freedom) the oscillator goes from the
initial ground state to the vacuum squeezed state \cite{D95}, whose properties are
well known. In particular, this state has a strongly oscillating photon
distribution function (abbreviated hereafter as PDF) $f(m)\equiv \langle m|\hat\rho| m\rangle$
(i.e., the probability to detect $m$ quanta in the state
described by the statistical operator $\hat\rho$):
\be
f(2m)= \frac{\langle n\rangle^m (2m)!}{(1+\langle n\rangle)^{m+1/2}(2^m m!)^2},
\quad f(2m+1)=0,
\label{pdf-sqz}
\ee
where $\langle n\rangle$ is the mean number of quanta.
Another important property of the vacuum squeezed state is a high degree of quadrature squeezing 
when $\langle n\rangle \gg 1$.

A quite different situation arises if the intermode interaction is
essential (as happens in the effectively one-dimensional Fabry--Perot cavity
\cite{DK92,DA99} or in the three- and two-dimensional cavities with accidental degeneracies
of the eigenmode spectra \cite{Croc1,DD01}) or coupling with a detector is strong
enough \cite{D95}. Then the degree of squeezing in each mode becomes much smaller
than in the vacuum squeezed state
and the oscillations of the PDF disappear. 

New features appear when
the dissipation becomes important. Just such a situation takes place
in the MIR experiment. Its main idea
is to simulate a motion of one of the cavity walls using an effective electron-hole `plasma mirror',
created periodically on the surface of a semiconductor slab (attached to the wall)
by illuminating it with a sequence of short laser pulses.
If the interval between pulses exceeds the recombination time of
carriers in the semiconductor, a highly conducting layer will
periodically appear and disappear on the surface of the semiconductor
film, thus simulating periodical displacements of the boundary.
In this way, rather big relative amplitudes of displacements and changes of the
fundamental cavity eigenfrequency, of the order of $10^{-3}$ or $10^{-2}$, can be
easily achieved using standard semiconductor plates having the thickness of the order
of $1\,$mm. This is is a great advantage over the schemes where real oscillations of the surface of the
cavity walls are excited, because in the latter case the relative amplitude of displacements cannot
exceed the value $10^{-8}$ due to tremendous internal stresses arising inside the material for
the frequencies of the order of a few GHz or higher.

Note that the thickness of the photo-excited conducting layer nearby the surface of the semiconductor slab is much smaller
than the thickness of the slab itself. It is determined mainly by the absorption coefficient of the laser radiation,
so it is about few micrometers or less, depending on the laser wavelength. Therefore laser pulses with the surface
energy density about few $\mu$J/cm$^{2}$ can create a highly conducting layer with the carrier concentration
exceeding $10^{17}\,$cm$^{-3}$, which gives rise to an almost maximal possible change of the cavity eigenfrequency 
for the given geometry \cite{DD-JPB}.
It is worth noting that although the thickness of the conducting layer is less than the skin depth, it gives the
same frequency shift as the conductor filling in all the slab. This interesting fact was explained and
verified experimentally in \cite{Pad-freq}.
On the other hand, the conductivity of the layer is not extremely high due to a moderate value of the mobility
in the available materials (such as highly doped GaAs), which is less than $1\,$m$^2$V$^{-1}$s$^{-1} $ \cite{Pad-Cas60}.
For this reason, effects of dissipation during the excitation-recombination process inside the layer
 cannot be neglected, since they can change the picture drastically \cite{DD-rev2,DD-JPB}. 
A model taking into account the dissipation was developed in \cite{DD-rev2},
and some of its consequences with respect to the photon statistics and PDF
were considered recently in \cite{DCas60,D09}. Here I analyze some
interesting special cases that were not considered in the previous papers.

\section{Exact formulas for the PDF and statistical moments}

I assume that the effects of dissipation can be taken into account by means of a
model based on the Heisenberg--Langevin operator equations of the form ($\hbar=1$)
\be
d\hat{x}/dt = \hat{p} -\gamma_x(t)\hat{x} +\hat{F}_x(t),
\label{Fx}
\ee
\be
d\hat{p}/dt = -\gamma_p(t)\hat{p} -\omega^2(t)\hat{x} +\hat{F}_p(t),
\label{Fp}
\ee
where $\hat{x}$ and $\hat{p}$ are dimensionless quadrature operators of the selected
mode of the EM field. These operators are
normalized in such a way that the mean number of photons equals 
${\cal N}=\frac12\langle \hat{p}^2 + \hat{x}^2 -1\rangle$.
Two noise operators 
$\hat{F}_x(t)$ and $\hat{F}_p(t)$ with zero mean values (commuting with $\hat{x}$ and $\hat{p}$)
are necessary to preserve the canonical commutator 
$\left[\hat{x}(t),\hat{p}(t)\right] = i$. 
These operators give a simplified description of 
 complicated processes inside a thin lossy dielectric
(semiconductor) slab attached to one of the cavity walls. In the phenomenological model used
in this paper, the net result of all those processes is encoded in the correlators of the
noise operators (the Markov approximation is assumed) 
\be
\langle \hat{F}_j(t) \hat{F}_k(t')\rangle =\delta(t-t')\chi_{jk}(t),
\qquad j,k = x,p.
\label{sigjk}
\ee
In principle, the functions $\chi_{jk}(t)$, as well as the functions $\gamma_x(t)$ and $\gamma_p(t)$,
 should be derived from some `microscopical' model
of electron--photon and electron--phonon interactions inside the semiconductor slab.  
In the phenomenological model considered here, these coefficients can only be  `guessed'.
The simplest possibility consistent with principles of quantum mechanics is as follows
\cite{DD-rev2,D09}:
\be
\gamma_x(t) = \gamma_p(t) =\gamma(t), 
\label{gam}
\ee
\be
\chi_{xp}=-\chi_{px} =i\gamma(t), \quad \chi_{xx} = \chi_{pp} =\gamma(t) G,
\label{choice}
\ee
where 
$ \gamma(t)$ is identified with the imaginary part
of the complex time-dependent eigenfrequency of the cavity
$\omega_c(t)=\omega(t) - i\gamma(t)$, which can be found
from the solution of the classical electrodynamical problem by
taking the {\em instantaneous\/} geometry and material properties (for example, by solving the Helmholtz equation with
the complex dielectric function $\vep(t)$ inside the slab and the set of cavity dimensions $\{L_j(t)\}$, 
where the time variable $t$ is considered as a {\em parameter\/}). 
The coefficient $G$ is related to the temperature of the reservoir $\Theta$ as
\be
G=1+2\langle n \rangle_{th} = \coth\left[{\hbar\omega_i}/{(2k_B \Theta) }\right]
\label{def-G}
\ee
(so that $\langle n \rangle_{th}$ is the mean number of quanta for the given cavity mode in the
thermodynamic equilibrium). 
The specific set of coefficients (\ref{gam}) and (\ref{choice}) has the following remarkable property: 
if the frequency $\omega$ does not depend on time, then
the second-order statistical moments 
$\langle \hat{x}^2\rangle$, $\langle \hat{p}^2\rangle$
and $\langle \hat{x}\hat{p} + \hat{p}\hat{x}\rangle$ tend to the well known equilibrium values
for an arbitrary positive function $\gamma(t)$ (without any corrections).
Some arguments {\em pro e contra\/} the choice of damping coefficients in
the form (\ref{gam}) can be found in \cite{D09}.

Equations (\ref{Fx}) and (\ref{Fp})  can be solved
explicitly for arbitrary time-dependent functions $\gamma_{x,p}(t)$,
$\omega(t)$ and $\hat{F}_{x,p}(t)$:
\be
\hat{x}(t) = \hat{x}_s(t) + \hat{X}(t), \qquad
\hat{p}(t) = \hat{p}_s(t) + \hat{P}(t).
\label{s-X}
\ee
The first terms are the solutions of homogeneous equations
(without the noise operators)
\be
\hat{x}_s(t) = e^{-\Gamma(t)}\left\{
\hat{x}_0 {\mbox Re}\left[\xi(t)\right]
-\hat{p}_0 {\mbox Im}\left[\xi(t)\right] \right\} ,
\label{xX}
\ee
\be
\hat{p}_s(t) = e^{-\Gamma(t)}\left\{
\hat{x}_0 {\mbox Re}\left[\dot\xi(t)\right]
-\hat{p}_0 {\mbox Im}\left[\dot\xi(t)\right] \right\},
\label{pP}
\ee
where $\hat{x}_0$ and $\hat{p}_0$ are the values of operators
at $t=0$ (taken as the initial instant) and
$
\Gamma(t)=\int_{0}^{t} \gamma(\tau)d\tau$.
Function $\xi(t)$ is a special solution to the classical oscillator equation 
\be
\ddot{\xi} +\omega^2(t) \xi=0,
\label{claseq}
\ee
 selected by
the initial condition $\xi(t)=\exp(-it)$
for $t\to-\infty$, which is equivalent to fixing the value of the Wronskian
\be
\xi\dot\xi^* -\dot\xi \xi^* =2i.
\label{vep}
\ee

The operators $\hat{X}(t)$ and $\hat{P}(t)$ 
represent the influence of the stochastic forces:
\be
\left(
\begin{array}{c}
\hat{X}(t)
\\
\hat{P}(t)
\end{array}
\right)
 = e^{-\Gamma(t)}
\int_{0}^t d\tau e^{\Gamma(\tau)}
{\cal A}(t;\tau)
\left(
\begin{array}{c}
\hat{F}_x(\tau)
\\
\hat{F}_p(\tau)
\end{array}
\right)
,
\label{XP-a}
\ee
where the $2\times2$ matrix 
\be
{\cal A}(t;\tau)=
\left(
\begin{array}{cc}
a_{x}^{x}(t;\tau) & a_{x}^{p}(t;\tau)
\\
a_{p}^{x}(t;\tau) & a_{p}^{p}(t;\tau)
\end{array}
\right)
\ee
consists of the following elements:
\be
a_{x}^{x}= {\mbox Im}\left[\xi(t)\dot\xi^*(\tau)\right], \quad
a_{x}^{p}= {\mbox Im}\left[\xi^*(t)\xi(\tau)\right],
\label{ax}
\ee
\be
a_{p}^{x}= {\mbox Im}\left[\dot\xi(t)\dot\xi^*(\tau)\right], \quad
a_{p}^{p}= {\mbox Im}\left[\dot\xi^*(t)\xi(\tau)\right].
\label{ap}
\ee

Combining equations (\ref{XP-a})-(\ref{ap}) with (\ref{sigjk})-(\ref{choice})
one can obtain the following expressions for
the second-order moments of  operators $\hat{P}(t)$ and $\hat{X}(t)$
[here $f_t \equiv f(t)$]: 
\be
\langle \hat{P}^2(t)\rangle = |\dot\xi_t|^2 J_{t}
-\mbox{Re}\left(\dot\xi_t^{*2}\tilde{J}_{t}\right),
\label{P2JY}
\ee
\be
\langle \hat{X}^2(t)\rangle = |\xi_t|^2 J_{t}
-\mbox{Re}\left(\xi_t^{*2}\tilde{J}_{t}\right),
\label{X2JY}
\ee
\be
\frac12\langle \hat{X} \hat{P} +\hat{P} \hat{X}\rangle_t = 
\mbox{Re}\left(\xi_t\dot\xi_t^{*} J_{t} -
\xi_t^{*}\dot\xi_t^{*}\tilde{J}_{t}\right),
\label{XPJY}
\ee
\be
J_{t} = 
\frac{G}{2} e^{-2\Gamma(t)}
\int_{0}^t d\tau e^{2\Gamma(\tau)}\gamma(\tau)
\left(|\xi_{\tau}|^2 + |\dot\xi_{\tau}|^2 \right)
,
\label{J}
\ee
\be
\tilde{J}_{t} = 
\frac{G}{2} e^{-2\Gamma(t)}
\int_{0}^t d\tau e^{2\Gamma(\tau)}\gamma(\tau)
\left(\xi_{\tau}^2 + \dot\xi_{\tau}^2 \right).
\label{tildeJ}
\ee

%\section{The mean number and number variance of quanta}
%\label{sub-mean}

The mean number of quanta can be written as
${\cal N}(t)= {\cal N}_s(t) + {\cal N}_r(t)$,
 where the first term
depends on the initial state (`signal'), while the second term is
determined by the interaction with the reservoir.
From (\ref{P2JY}) and (\ref{X2JY}) one obtains
\be
{\cal N}_r(t) =  E_{t}J_{t} 
 -\mbox{Re}\left(\tilde{E}_{t}^* \tilde{J}_{t} \right),
\label{dN}
\ee
%where
\be
E_{t}= \frac12\left(|\xi_t|^2 + |\dot\xi_t|^2\right), 
\quad
\tilde{E}_{t}= \frac12\left( \xi^2_{t} + \dot\xi^2_{t}\right).
\label{Etau}
\ee
For the initial thermal state characterized by the parameter $G_0$
(which can be different from $G$) one has
\be
{\cal N}^{(th)}_s(t)= \frac12\left\{G_0 e^{-2\Gamma(t)} E(t)
-1 \right\}.
\label{Nthermgen1}
\ee
One should remember that
formulas  (\ref{dN}) and (\ref{Nthermgen1}) 
make sense for sufficiently big values of time $t$, when the laser pulses
have been switched off and the recombination processes have been over, so that the
normalized frequency $\omega(t)$ returns to its initial unit value
(because the photon number operator
$\hat{\cal N}=\frac12\left( \hat{p}^2 + \hat{x}^2 -1\right)$
is defined with respect to the initial geometry of the cavity, coinciding with the final one).

I consider here only the special (although the most realistic) case of initial
{\em thermal\/}  states of the field.
It is well known
\cite{Lax,Haken} that the description of open quantum systems by
means of the Heisenberg--Langevin equations with delta-correlated
stochastic force operators is equivalent to the description in the
Schr\"odinger picture by means of the master equation for the
statistical operator. In the case of {\em linear\/} operator equations of
motion, such as equations (\ref{Fx}) and (\ref{Fp}), the corresponding
master equations contain only {\em quadratic\/} terms (various products
of {\em two\/} operators $\hat{p}$ and $\hat{x}$) \cite{Lax,Haken,167,JRLR95}).
Consequently, any initial {\em Gaussian\/} state (whose special case is the thermal state) 
remains Gaussian in the process of evolution. 
For thermal states, mean values of the first order moments are equal to zero,
%$\langle \hat{ x}\rangle = \langle \hat{ p}\rangle =0$,
and this property is preserved in the process of evolution governed by 
equations (\ref{Fx}) and (\ref{Fp}). Then all statistical properties of the
single mode are determined completely by the variances of the quadrature
operators $\sigma_{xx}$, $\sigma_{pp}$ and by their covariance $\sigma_{xp}=\sigma_{px}$ 
(in the case involved, 
$\sigma_{ab}= \frac12\langle \hat{a}\hat{b} + \hat{b}\hat{a}\rangle$).
Using equations (\ref{xX}) and (\ref{pP}) one can verify that the time-dependent
(co)variances can be obtained from formulas
 (\ref{P2JY})-(\ref{XPJY}) by means of the replacement
$J_t \longrightarrow J_t + \frac12 G_0\exp(-2\Gamma_t)$.

The photon distribution function $f(m)\equiv \langle m|\hat\rho| m\rangle$
of the Gaussian states was found long ago \cite{AgAd,Chat,Mar,MarMar,1mod,book}.
For zero mean values $\langle \hat{ x}\rangle = \langle \hat{ p}\rangle =0$
it can be expressed in terms of the Legendre polynomials 
\be
f(m)=
\frac{2D_{-}^{m/2}}{D_{+}^{(m+1)/2}}
P_m\left(\frac{4\Delta -1}
{\sqrt{D_{+}D_{-}}}\right)
\label{dist0}
\ee
where
\be
D_{\pm} = 1+4\Delta \pm 2\tau,
\label{Dpm}
\ee
\be
\tau = \sigma_{xx}+ \sigma_{pp} \equiv 1 +2{\cal N},
\label{def-tau}
\ee
\be
\Delta = \sigma_{xx}\sigma_{pp} -\sigma_{px}^2 \ge 1/4 
\label{def-kap}
\ee
(the last inequality is the Schr\"odinger--Robertson uncertainty relation).
For the Gaussian quantum states one can write
 $\Delta= 1/(4\mu^2)$, where
the quantity $\mu$ is the {\em quantum purity\/} of the state:
$\mu = \mbox{Tr}(\hat\rho^2)$. 
For the pure Gaussian quantum states (in the absence of dissipation) $\Delta \equiv 1/4$,
and formula (\ref{dist0}) goes directly to (\ref{pdf-sqz}).

The explicit expression for the time-dependent coefficient $\Delta$
is as follows:
\be
\Delta = \left(J + \frac{G_0}{2} e^{-2\Gamma}\right)^2 -  |\tilde{J}|^2. 
\label{Delta-JJ}
\ee

If the
functions $\omega(t)=\omega_0[1+\chi(t)]$ and $\gamma(t)$ have the form
of {\em periodical\/} pulses 
separated by intervals of time with $\omega=\omega_0=const$
and $\gamma=0$ (this means that the quality factor of the cavity is
supposed to be high enough), then
the following formulas can be
obtained for the quantities
$E_n\equiv E(nt)$, $\tilde{E}_n\equiv \tilde{E}(nt)$, 
 $J_n\equiv J(nt)$ and $\tilde{J}_n\equiv \tilde{J}(nt)$ 
after $n$ periods ($n$ pulses of laser irradiation in the case of DCE)
under the realistic conditions  $|\chi(t)| \ll 1$ and $\gamma(t) \ll 1$ \cite{DD-rev2,D09}:
\be
E_n= \cosh(2n\nu), \quad
\tilde{E}_n= \sinh(2n\nu) e^{i\beta},
\label{EkEkT}
\ee
\be
J_n = A_n^{(+)} + A_n^{(-)}, \quad \tilde{J}_n = e^{i\beta}
\left(A_n^{(+)} - A_n^{(-)}\right),
\label{JAB}
\ee
\be
A_n^{(\pm)}= \frac{G\Lambda}{4(\Lambda \pm \nu)}\left(e^{\pm 2n\nu} - e^{-2n\Lambda}\right),
\label{An}
\ee
where $\beta$ is some insignificant constant phase, %(which is not important for the following analysis),
\be
\nu =
\left| \int_{t_i}^{t_f} \omega_0
{\chi}(t) e^{-2i\omega_0 t} dt \right|
, \qquad
\Lambda =\int_{t_i}^{t_f} \gamma(\tau)d\tau.
\label{Lam}
\ee
Here $t_i$ and $t_f$ are the initial and final moments of each pulse. 
It is taken into account that $\Lambda, \nu \ll 1$.
The mean number of photons grows exponentially under the conditions
$2n\nu \gg 1$ and $\nu >\Lambda $:
\be
{\cal N}_{n} = \frac14 e^{2n(\nu -\Lambda)}\left( G_0 +\frac{G\Lambda}{\nu - \Lambda}\right)
+{\cal O}(1).
\label{Ntotas}
\ee
Coefficient $G_0$ is given by formula (\ref{def-G}), but with $\Theta$ replaced
by the initial temperature of the field mode $\Theta_0$ 
(which can be made different from $\Theta$).
Under the same conditions formula (\ref{Delta-JJ}) takes the form
\be
\Delta_n = {\cal N}_n \frac{G\Lambda}{\nu + \Lambda} +{\cal O}(1).
\label{Del-n}
\ee
Note that the ratio $\Delta_n / {\cal N}_n$ in this limit does not depend
on the coefficient $G_0$ (the initial temperature of the field mode).
Formulas (\ref{EkEkT})-(\ref{Del-n}) hold provided the periodicity of pulses $T$ is adjusted
to the initial period of oscillations of the field mode $T_0=2\pi/\omega_0$
as follows ($m=1,2,\ldots$):
\be
T=  \frac12 T_0\left(m - {\varphi}/{\pi}\right),
\quad
\varphi = - \omega_0 \int_{t_i}^{t_f}{\chi}(t)  dt.
\label{def-delT}
\ee
A small shift $\varphi$ of the resonance periodicity of pulses arises
if the profile of pulses is asymmetrical (namely this situation takes place
in reality).

\section{Asymptotical formulas for the photon statistics and squeezing}
\label{sec-photstat}

Formula (\ref{dist0}) is exact. However, it is not very
convenient for calculations in the case of DCE if
the number of created photons
%created due to the parametric resonance
is big (say, $m \sim {\cal N} > 1000$; otherwise the effect cannot be confirmed
with certainty at the existing experimental level due to the noise in the measurement channel).
Therefore asymptotical forms of exact formulas for $m\gg 1$
can be more useful.
Note that the argument of the Legendre polynomial in (\ref{dist0})
is always outside the interval $(-1,1)$,
being equal to unity only for thermal states with $\tau=2\sqrt{\Delta}$. 
For this reason it is convenient to use the asymptotical formula \cite{Olver}
\begin{equation}
P_m(\cosh\xi )\approx
\left(\frac {\xi}{\sinh\xi}\right)^{1/2}
I_0\left(\left[m+1/2\right]\xi\right),
\label{as-Olv}
\end{equation}
where $I_0(z)$ is the modified Bessel function.
Formula (\ref{dist0}) shows that the behavior of the PDF depends on the sign of the coefficient $D_{-}$,
i.e., on the ratio $2\Delta/\tau \approx \Delta_n/{\cal N}_n$. If this ratio exceeds the unit value,
then the argument of the Legendre polynomial is real (and bigger than unity).
In this case, using the known asymptotical formula
$I_0(x) \approx (2\pi x)^{-1/2}\exp(x)$ (if $x\gg 1$) and making some further simplifications,
one can arrive (under the conditions ${\cal N} \gg 1$ and $m\gg 1$) at the simple formula \cite{DCas60,D09}
\be
f(m) 
\approx \frac{\exp[-(m+1/2)/(2{\cal N})]}{\sqrt{2\pi{\cal N}(m+1/2)}}.
\label{fm-fin}
\ee
Here I consider in detail the case when $D_{-} <0$ or $\Delta_n/{\cal N}_n <1$.
Then formula (\ref{as-Olv}) is still valid, but it needs some transformations
because the parameter $\xi$ becomes complex.
It is clear from formula (\ref{dist0}) that the real positive value $f(m)$ does not
depend on the choice of sign of the square root function $\sqrt{D_{-}}= \pm i\sqrt{|D_{-}|}$,
provided this sign is maintaned, both inside the argument of the Legendre polynomial
and in the coefficient in front of this polynomial.
Choosing for definiteness the branch $\sqrt{D_{-}}=i\sqrt{|D_{-}|}$ it is convenient to write
\[
\xi = \tilde\xi -i\pi/2 = -i\eta, \quad
\sinh(\tilde\xi) = \frac{4\Delta -1}{\sqrt{D_{+}|D_{-}|}},
\]
so that $\tilde\xi$ is real positive number. Using the relation
$I_0(-i\eta)=J_0(\eta)$ in (\ref{as-Olv}), one can represent (\ref{dist0}) as
\[
f(m) \approx i^m \left(\frac{2\eta}{r}\right)^{1/2}\left|\frac{D_-}{D_+}\right|^{\frac{m}2 +\frac14}
J_0\left(\left[m+1/2\right]\eta\right),
\]
where $r=\sqrt{\tau^2 -4\Delta}$.
Since $\arg(\eta)<\pi/2$, the asymptotical formula
$J_0(z) \sim \sqrt{\frac{2}{\pi z}}\cos(z-\pi/4)$ can be used, leading to the formula
\[
f(m) \approx  \frac{2 i^m \cos\left(m\pi/2 +i\left[m+1/2\right]\tilde\xi\right)}
{\sqrt{\pi(m+1/2)r}}\left|\frac{D_-}{D_+}\right|^{\frac{m}2 +\frac14}
.
\]
Replacing the cosine function by the sum of two exponentials, one obtains
\beqn
f(m) &\approx&  \frac{1}{\sqrt{\pi(m+1/2)r}}\left[
\left(\frac{2r +\delta}{D_+}\right)^{m +\frac12} \right. \nonumber
\\ && \left.
 +(-1)^m \left(\frac{|D_-|}{2r +\delta}\right)^{m +\frac12}\right],
\label{fDneg}
\eeqn
where $\delta=4\Delta-1$.
The right-hand side of (\ref{fDneg}) is real and positive.
 Using the relation $2\Delta =b\tau +c$
for $\tau \gg 1$ [with $b=G\Lambda/(\nu +\Lambda)$ according to equation (\ref{Del-n})],
one can verify the relations
\be
\frac{2r +\delta}{D_+} = 1 -\frac1{\tau} +{\cal O}\left(\tau^{-2}\right), 
\label{fracD+}
\ee
\be
\frac{|D_-|}{2r +\delta} = \frac{1-b}{1+b}\left[
1 - \frac{b^2 +2c}{\tau(1-b^2)}+{\cal O}\left(\tau^{-2}\right)\right].
\label{fracD-}
\ee
Note that the coefficient at $\tau^{-1}$ in (\ref{fracD+}) does not contain
the coefficients $b$ and $c$. Obviously, the expansion (\ref{fracD-}) is valid
provided $1-b$ is not too small.
Formulas (\ref{fDneg}) and (\ref{fracD-}) clearly show that oscillations of
the PDF are actually negligible (especially for $m \sim {\cal N}$),
 unless $b\ll 1$ (i.e., in the case of extremely
small dissipation). In the latter case the coefficient $c$ can be expressed
in terms of the initial purity which, in turn, is related to the factor $G_0$
as $\mu = 1/G_0$, so that $c=G_0^2/2$. Using the approximate formula
$(1-x)^m \approx \exp(-mx)$, which holds for $x \ll 1 $ and $mx^2 \ll 1$,
one can rewrite (\ref{fDneg}) in the case of small dissipation as (replacing
$r$ by $\tau$)
\beqn
f(m) &\approx& \left[\pi\tau(m+1/2)\right]^{-1/2}\left\{
\exp\left[-(m+1/2)/\tau\right] 
\right. \nonumber \\ && \left.
+ (-1)^m \exp\left[-(m+1/2)G_0^2/\tau\right]\right\}.
\label{fmosc} 
\eeqn
Formula (\ref{fmosc}) is valid under the conditions $\tau\approx 2{\cal N} \gg 1$, 
$1 \ll m \ll \tau^2$ and $G\Lambda/\nu \ll 1$. Even in this case the oscillations
of the PDF can be noticed only if $G_0 \sim 1$, i.e., for low temperature
initial thermal states. In particular, (\ref{fmosc}) coincides with the asymptotical
form of the ideal PDF of the vacuum squeezed state (\ref{pdf-sqz}) if $G_0=1$ and
$\tau\approx 2\langle n\rangle \gg 1$.
In the conditions of the MIR experiment the ratio $\Lambda/\nu$
exceeds $1/2$ \cite{DCas60}, which means that the oscillating term in (\ref{fDneg})
can be neglected even for $G=1$ (zero temperature of the cavity walls), so that
formula (\ref{fm-fin}) can be used for any values of parameters $G_0$ and $G$
(under the conditions ${\cal N} \gg 1$ and $m\gg 1$).

Using the Euler--MacLaurin summation formula, one can verify that 
the distribution function (\ref{fm-fin}) has the correct normalization 
with an accuracy ${\cal O}(\tau^{-1/2})$:
\beqn
&&\sum_{m=0}^{\infty} f(m) \approx  \int_0^{\infty} f(m) dm
+ {\cal O}[f(0)] 
\\ &&
\approx \int_0^{\infty} \frac{\exp(-x/\tau) }
{\sqrt{\pi\tau x}} dx
 + {\cal O}(\tau^{-1/2}) = 1 + {\cal O}(\tau^{-1/2}).
\label{norm}
\eeqn
The oscillating terms in (\ref{fDneg}) or (\ref{fmosc}) do not influence the normalization,
because they give corrections of a higher order. For example, combining the nearest
positive and negative terms with $m=2k$ and $m=2k+1$ and applying the Euler--MacLaurin summation formula
to the sum over these positive pairs (which contains now only slowly varying positive terms),
one can see that the correction equals approximately $G_0/(2\tau)$.

With the same accuracy as in (\ref{norm}), the moments of the distribution function can
be calculated as
\beqn
\langle m^k \rangle &\equiv & \sum_{m=0}^{\infty} m^k f(m)
\approx \int_0^{\infty} x^k \frac{\exp(-x/\tau) }
{\sqrt{\pi\tau x}} dx 
%\nonumber 
\\
&=& \tau^k \frac{(2k-1)!!}{2^k} \approx
{\cal N}^k (2k-1)!! \,.
\label{m-k}
\eeqn
For $k=2$ formula (\ref{m-k}) leads to the following formula for the
variance of the number of created quanta:
$\sigma_N = \langle m^2 \rangle\ -\langle m \rangle^2 \approx 2{\cal N}^2$.
It means that the field mode goes asymptotically to the so called
`superchaotic' \cite{McNeil75,Sot81} quantum state, whose statistics is essentially
 different from the statistics of the initial thermal state, characterized by
formula $\sigma_N ={\cal N}({\cal N}+1) \approx {\cal N}^2$.
The same result can be obtained from the general formula for the variance
of the number of quanta in the Gaussian quantum states \cite{1mod}
$\sigma_N = \tau^2/2 -\Delta -1/4$ (if $\langle \hat{ x}\rangle = \langle \hat{ p}\rangle =0$).

It is interesting that the asymptotical value of the ratio $2\Delta/\tau =b$ in the case
concerned coincides with the {\em invariant squeezing coefficient}. This coefficient
can be introduced in the following way.
Obviously, instantaneous values of variances $\sigma_{xx}$, $\sigma_{pp}$ and
$\sigma_{xp}$ cannot serve as true measures of squeezing,
since they depend on time in the course of the free evolution of
the oscillator.
For example (in dimensionless units with $\omega_0=1$),
\be
\sigma_{xx}(t)=\sigma_{xx}^{(0)}\cos^2(t) +\sigma_{pp}^{(0)}\sin^2(t)
+\sigma_{xp}^{(0)}\sin(2t),
\label{sig-t}
\ee
 and it can happen that both variances
$\sigma_{xx}$ and $\sigma_{pp}$ are large, but nonetheless the state
is highly squeezed due to the large nonzero covariance $\sigma_{xp}$.
It is reasonable to introduce some invariant characteristics which do not
depend on time in the course of free evolution (or on phase angle in the
definition of the field quadrature as
$\hat{E}(\varphi)=\left[\hat{a}\exp(-i\varphi)+\hat{a}^{\dagger}
\exp(i\varphi)\right]/\sqrt{2}$). 
I define the invariant squeezing coefficient $S$ as the ratio
of the minimal value of the variance $\sigma_{xx}(t)$ [as a function of time (\ref{sig-t})]
to the dimensionless variance $1/2$ in the vacuum state. Then straightforward calculations
give the formula (similar results can be found in \cite{book,LuPeHr,Loud89,GSQ})
\be
S = \frac{4\Delta}{\tau + \sqrt{\tau^2 -4\Delta}},
\label{S}
\ee
so that, indeed, $S=b=G\Lambda/(\nu +\Lambda)$ in the case discussed.
In contradistinction to the ideal case ($\gamma=0$), when $\Delta=const$, so that an arbitrarily
small invariant squeezing coefficient $S_{id} \approx G_0^2/(4{\cal N})$ can be obtained,
in the presence of dissipation the squeezing coefficient goes to the constant value which does
not depend on the initial temperature of the mode but depends on the temperature of the
reservoir (the cavity walls). 

\section{Conclusion}

The main results of the paper are as follows. New formulas (\ref{fDneg}) and (\ref{fmosc})
show how the increase of dissipation and temperature transform the strongly oscillating photon
distribution function (\ref{pdf-sqz}) of the ideal squeezed vacuum state to the smooth
asymptotical distribution (\ref{fm-fin}). Namely this smooth distribution is expected to be
observed in the MIR experiment due to strong dissipation. 
Another new result is the explicit demonstration of the correlation between the existence of
squeezing (when the invariant squeezing coefficient $S$ is less than unity) 
and the oscillations of the PDF in the case concerned. In particular, no squeezing
of the fundamental field mode is expected in the MIR experiment.

\section*{Acknowledgments}

A partial support of the Brazilian agency CNPq is acknowledged.

\end{document}